# Communication and Knowledge:
# How is the knowledge base of an economy constructed?

Loet Leydesdorff
University of Amsterdam, Science & Technology Dynamics
Amsterdam School of Communications Research (ASCoR),
Kloveniersburgwal 48, 1012 CX Amsterdam, The Netherlands
http://www.leydesdorff.net ; loet@leydesdorff.net

## Abstract

The competitive advantages in a knowledge-based economy can no longer be attributed to single nodes in the network. Political economies are increasingly reshaped by knowledge-based developments that upset market equilibria and institutional arrangements. The network coordinates the subdynamics of (i) wealth production, (ii) organized novelty production, and (iii) private appropriation versus public control. The interaction terms generate a complex dynamics which cannot be expected to contain central coordination. However, the knowledge infrastructure of systems of innovations can be measured, for example, in terms of university-industry-government relations. The mutual information in these three dimensions indicates the globalization of the knowledge base. Patent statistics and data from the Internet are compared in terms of this indicator.

## Introduction

Whereas organizations and institutions can be identified as observable units of analysis or nodes in a network, knowledge develops operationally in terms of reconstructions at the level of the links. Knowledge flows both within organizations and across institutional boundaries. In order to study organized knowledge production, therefore, one first has to distinguish analytically between the intellectual and the institutional organization of knowledge production systems (Whitley, 1984). The intellectual organization functions over time in terms of communications and their codifications, whereas the institutional organization provides structural coordination at each moment in time (Luhmann, 1984; Cowan & Foray, 1997). The 'knowledge base' can thus be considered as an overlay of mutual expectations that feedback on the institutional arrangements among the knowledge organizers (Leydesdorff & Etzkowitz, 2001).

Network arrangements provide the background for knowledge flows (Castells, 1996; David & Foray, 1994). In a knowledge-based economy the institutional arrangements among knowledge organizers (e.g., universities, industries, and governmental agencies) can become a necessary condition for both producing and retaining wealth from knowledge (e.g., Popper & Wagner, 2002; Steinmueller, 2002). Because of the potential overlap in networks at different levels, one can no longer expect the organization of knowledge to be contained within a single organization.

In other words, the 'knowledge base' of an economy generates a dynamics orthogonal to that of the knowledge infrastructure of a political economy. The latter provides arrangements that stabilize the market system at each moment, while knowledge flows through these networks in fluxes with different speeds. The interaction among the flows puts pressure on the previously established boundaries. For example, pharmaceutical corporations can nowadays no longer carry the costs of biotechnological innovations without relying on knowledge networks (Owen-Smith *et al.*, 2002). Corporate boundaries increasingly function as mechanisms for the appropriation and shielding of competitive advantages from the knowledge fluxes through the networks.

*Knowledge-based innovations* change the interfaces between supply-side agencies producing novelty (e.g., R&D) and—market or non-market—selection environments. In this process the relevance of previously defined boundaries can be redefined. When the new boundaries become functional for the reproduction of the systems, new retention mechanisms can also become institutionalized. Knowledge-based innovations can thus be considered as the evolutionary operators that change the network structures in which they are reflexively generated (Fujigaki, 1998; Leydesdorff, 2001a).

### The knowledge-based economy

In the period before the oil crises of the 1970s, that is, in the decades after World War II, social functions were largely organized into institutions on a one-to-one basis (Merton, 1942; Bush, 1945). The global effects of the oil crises made clear that advanced industrial nations could outcompete low-wage countries only on the basis of the systematic exploitation of their respective knowledge bases (e.g., Nelson & Winter, 1977, 1982; Freeman, 1982). Collaboration across institutional boundaries, however, implies transaction costs (Williamson, 1975). The new relations may also generate longer-term revenues and synergies (e.g., Faulkner & Senker, 1995).

The transaction costs can be considered as a macro-investment in establishing new structures of collaboration and competition at the national level. Thus, a dynamic view of a knowledge-based system could be generated in which institutional agents have continuously to trade-off among optimizations using a variety of criteria (Galbraith, 1967). The trade-off between short and long-term costs and benefits brought governments into play in the interaction between R&D and the economy (OECD, 1971). Technological innovation policies were increasingly formulated (OECD, 1980).

During the 1980s, the techno-sciences like biotechnology, information technologies, and new materials rapidly became the top priorities for stimulation policies in the advanced industrial countries. Because these 'platform sciences' (Langford & Langford, 2001) are based on rearrangements across disciplinary lines—that is, recombination at the intellectual level—competitive advantages through synergies at this level are to be exploited for economic development at the institutional level (Andersen, 1994; Leydesdorff & Gauthier, 1996).[1] The reorganization and stimulation of university-industry relations at the institutional level thus became a second point of attention for S&T policy makers (Rothwell & Zegveld, 1981;

---

[1] Previous attempts at a more direct mission-oriented steering of the sciences had at that time been evaluated as less successful (Van den Daele, Krohn, & Weingart, 1977; Studer & Chubin, 1980).

OECD, 1988). Why had some countries been more successful than others in exploiting their knowledge-base (Hauff & Scharpf, 1975; Irvine & Martin, 1984)? Why had within countries certain sectors (e.g., chemistry, aircraft) been more successful than others in maintaining knowledge-intensive relations (Nelson, 1982)? Could lessons be learned from best practices across sectors and might such practices be transferable from one national context to another?

In the U.S.A. the national system experimented with granting universities the right to patents on the basis of federal funding (the Bayh-Dole Act of 1980), and systematic efforts were made to raise the level of knowledge-intensity within industry both at the level of the states and by stimulation programs at the level of the federal government (Etzkowitz, 1994; Spencer, 1997). Universities became increasingly players in the patent system of the U.S.A. (Henderson *et al.*, 1998) Thus, their role as systemic knowledge organizers in innovation networks became increasingly important (Etzkowitz & Leydesdorff, 2000; Owen-Smith *et al.*, 2002).

Figure 1 exhibits the percentage of patents that can be retrieved using the word 'university' as a search term in the database of the U.S. Patent and Trade Office (at http://www.uspto.gov/ ). The second curve—of the percentage of universities that can be retrieved using 'university' as a search term among the assignees of patents—shows even more clearly that the effect of the Bayh-Dole Act began to peak in 1997.

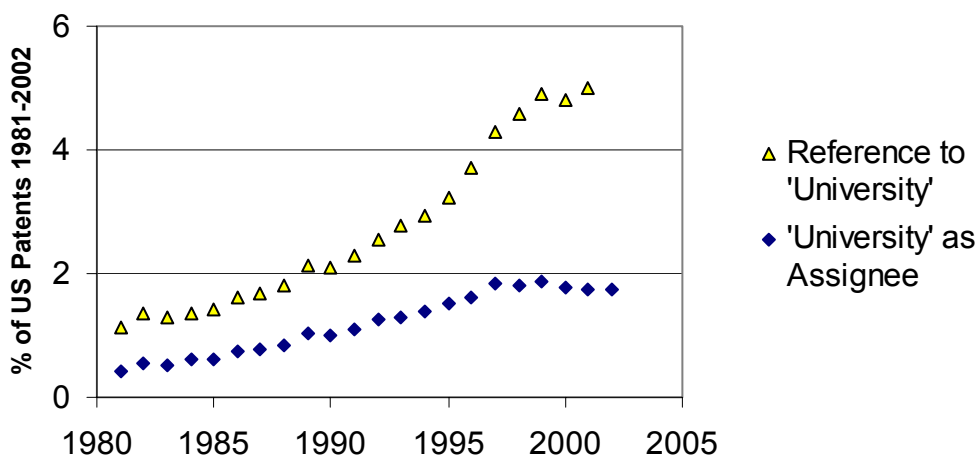

**Figure 1**
Percentage of U.S. Patents (i) with a reference to the word 'university' and (ii) a 'university' among the assignees

In the period between 1981 and 1997, universities have thus been enrolled as new players in the patenting domain (Henderson *et al.*, 1998; Sampat *et al.*, 2003). But what does this indicator teach us with respect to the role of academic research in innovation processes (Rosenberg & Nelson, 1994; Cohen *et al.*, 2002)? Whereas this role can be analyzed historically for innovations on a case-by-case basis, a specification of the relevant system(s) of innovations is needed to determine this role at the aggregate level.

# The delineation of systems of innovation

The definition of a system of innovations in terms of nations, sectors (Pavitt, 1984), technologies (Carlsson, 2002; Carlsson & Stankiewicz, 1991), regions (Braczyk *et al.*, 1998), etc., brings players other than the traditional ones into view. From the mid-1980s onwards, for example, the European Union has developed a series of Framework Programs containing new policies for science, technology, and innovation. Both trans-national cooperation and cooperation across sectors have been systematically stimulated. Within the emerging context of the European system, regions have tried deliberately to promote their positions as a relevant level for the systematic development of the knowledge infrastructure (Leydesdorff *et al.*, 2002)

Has a European system of innovations emerged in relation to the underlying national systems? Have regions (e.g., Catalonia, Flanders, etc.) been successful in establishing their own specific systems of innovation (Riba-Vilanova & Leydesdorff, 2001)? Have sectors (e.g., ICT) been developed using patterns of innovation which differ from those established in the previous cycles of industrial development (Barras, 1990)? How can systems of knowledge-based innovation be delineated and assessed if they cross national boundaries?

These questions became ever more pressing during the 1990s when the Internet emerged. South- and East-Asian countries seemed initially better equipped for moving ahead in the new e-environment given their specific mixes of human resources, flexibilities in industrial structures, and prevailing knowledge infrastructures. How should the previously advanced industrial countries react? Is it sufficient to stimulate ongoing processes of global change locally or should policy frameworks be proposed that enable new partnerships to be developed at the global level? Which criteria for the optimization should then be used (e.g., national, transnational, sectoral)? In other words, the stage was set for a deep reformulation of the very problem of science and technology policy-making in the first half of the 1990s.

# Science and technology policies in the 1990s

A redefinition of the problem of science and technology policies became urgent as the Internet signaled its future economic success in the first half of the 1990s. The additional dimension of global communication could be envisaged as changing the phase space of possibilities for international collaboration in science and technology, international trade, and international relations. Structural adjustments of existing arrangements were likely to gain further momentum (Freeman & Perez, 1988).

Gibbons *et al.* (1994) suggested making a distinction between 'Mode 2' and 'Mode 1' types of the production of scientific knowledge. Whereas 'Mode 1' refers to the traditional shape, largely confined within institutional settings, 'Mode 2' would be communication-driven. Knowledge can then be considered as a codification of communication. A scientific communication can be contained within an institution or even within an individual agent as 'tacit knowledge' and/or it can be 'published' and then brought into circulation.

These dimensions of public and private communication of *knowledge* resonate with and disturb the established public/private *arrangements* between industries and governments in the political economy. The knowledge component adds a reflexive dynamic to the so-called

'differential productivity growth puzzle' between various sectors in the economy (Nelson & Winter, 1975). Existing trade-offs between public control and the private appropriation of competitive advantages can be expected to be increasingly upset when innovations are systematically organized and stimulated (Nelson & Winter, 1977). New regulations (and perhaps new regulatory regimes) are needed when knowledge-based technologies restructure the sectoral organization (Callon, 1998). During the 1990s, increased knowledge-intensity became thus a driver to the reform of political economies.

In a number of papers, Henry Etzkowitz and I have proposed a neo-evolutionary model of a 'triple helix of university-industry-government relations' for these knowledge-based transformations of political economies (e.g., Etzkowitz & Leydesdorff, 1997 and 2000; Leydesdorff & Etzkowitz, 1998). As noted, three functions have to be fulfilled within a system of innovations: wealth generation in the economy, novelty and innovation production that upset the equilibrium seeking mechanisms in market systems, and public control versus private appropriation at the interfaces between economic systems of exchange and organized novelty production. The knowledge infrastructure of university-industry-government relations (e.g., at the level of a nation state) can be considered as a specific retention mechanism among these three subdynamics. These institutional arrangements and their trajectories are under pressure from global developments that function as a next-order regime (Dosi, 1982).

Advanced industrial states have historically generated 'national systems of innovation' during the past century or so based on the geographical proximity of the various subdynamics (Freeman, 1988; Lundvall, 1988, 1992; Nelson, 1993). The innovative knowledge flows, however, span boundaries and thus generate new types of competition at the global level (Krugman, 1996). In the Triple Helix model, this selection pressure is represented as an overlay of communications among the institutional agencies which have hitherto carried the knowledge infrastructure: industry, academia, and government. Each of these institutions is organized along international dimensions as well. At the level of the overlay of expectations, one can entertain and recombine possibilities other than those that have been realized hitherto. Thus, the linkages provide the carrying agencies with access to the knowledge-based system of coordination.

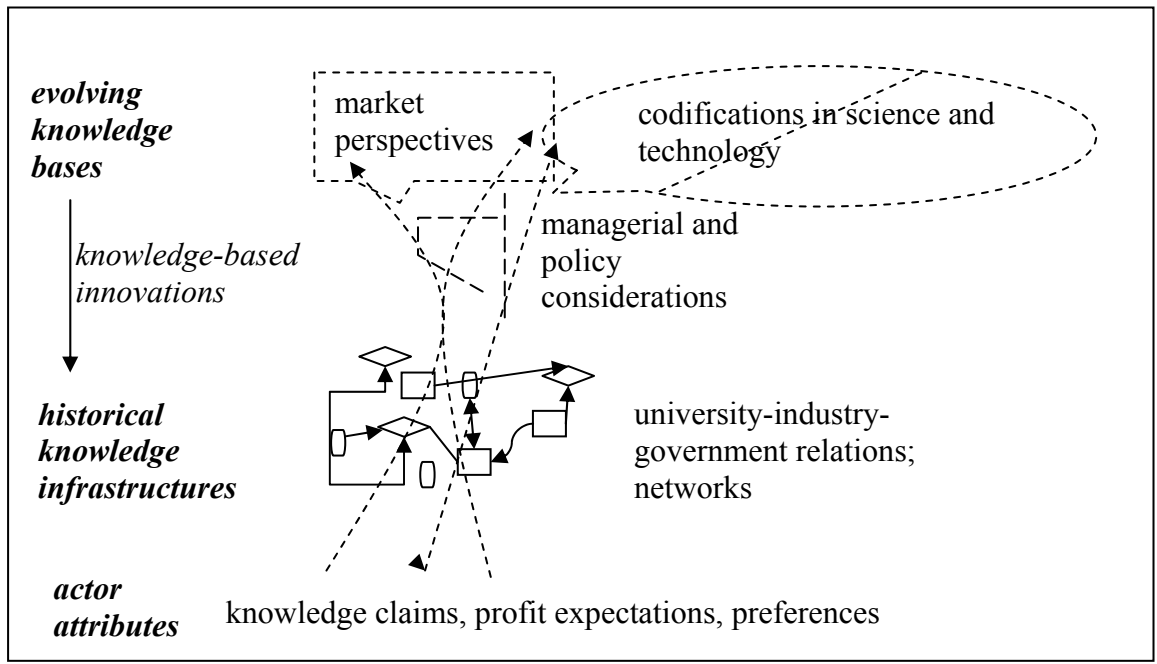

**Figure 2:** The evolving knowledge-base feeds back on the historical knowledge-infrastructure.

As the relative weights in the networks change by ongoing processes of collaboration, appropriation, and competition, new balances and unbalances can be expected to generate feedbacks in the knowledge infrastructure at other ends. For example, trajectories can be formed historically at interfaces when technologies are 'locked-in' within industries (e.g., the QWERTY keyboard or VHS tapes; cf. David, 1985; Arthur, 1989) or—similarly but between different subdynamics—when specific scientific expertise and government policies begin to co-evolve as they sometimes do in the energy and the health sectors (Elzinga, 1985; McKelvey, 1996). The state and industry can also become 'locked-in' as in the former Soviet Union.

Co-evolutions between two subdynamics continuously generate stabilities between counteracting mechanisms in processes of mutual shaping, whereas a third subdynamic potentially dissolves previous arrangements at a global level. The interacting subdynamics thus shape trajectories *and* regimes endogenously (Nelson, 1994; Leydesdorff & Van den Besselaar, 1998). Policies then have to vary according to which 'lock-ins' can be expected to prevail, and whether and how they can be disturbed. For example, the market mechanism can be expected to reintroduce flexibilities in the case of a bureaucratic lock-in, whereas in the case of a technological lock-in government interventions may be needed to break monopolistic tendencies. Thus, the policies become increasingly a variable *dependent* on the evolutionary assessment of the knowledge-based system.

## Localizable trajectories and globalized regimes

While the systems under study operate dynamically, knowledge flows between systems can temporarily be stabilized and further developed along the historical trajectories of institutions that have served the developments hitherto. For example, the well-organized niches of nation states can be considered as providing the stability that is necessary for accessing globalized—i.e., meta-stabilized—regimes (Luhmann, 2000, at p. 396). A global regime can be expected to emerge from closer interactions between hitherto relatively separate subsystems (Leydesdorff & Scharnhorst, 2003). The regime, however, contains a codification of the interactions among differently coded communication systems (e.g., the economy, science, and policy-making).

The emerging configurations of mutual expectations can be expected to change the selection pressure on the institutional arrangements that were shaped historically. The networks supporting the exchange have then to be restructured (Freeman & Perez, 1988). Older arrangements can be expected to survive, but they may change in terms of their functions as well as their forms (Frenken, 2000; Kauffman, 2000). However, one should not reify the 'global level agency' as a metabiology or a supersystem. The various systems of expectations interact and produce an overlay of global expectations *within* the network. The overlay globalizes the system by making representations available beyond the ones which could already be envisaged from the various (subsystemic) perspectives.

The envisaged recombinations can be attributed to a next-order or 'global' system, but this evolution remains an internal dynamics that is added to the system as its globalization. This globalization can be entertained reflexively and thus enrich the system. It provides a future-oriented knowledge-base that innovates the historically shaped structures with hindsight. The innovativeness is based on inventing new codifications reflexively by recombining perspectives at interfaces (e.g., between R&D and market perspectives).

Thus, the dynamics of knowledge systematically organized as science and technology induce a reflexive turn in social systems and, therefore, in the study of social systems. The 'reflexive turn' in science & technology studies (Woolgar & Ashmore, 1988) first implied that the idea of a single and universalistic yardstick—as sought in the philosophy of science (e.g., Popper, [1935] 1959)—had to be given up in favour of codes that can continuously be recombined and reconstructed. Unlike universal standards, however, asymmetry can be expected to prevail in exchange relations (Gilbert & Mulkay, 1984). The systems are able to exchange because they have different substances in stock.

For example, the political system is interested in results from the science system that inform decision making and legitimate policies, but without being burdened with the overwhelming uncertainties that are intrinsic to scientific inference (Beck, 1986). Within the science systems these uncertainties potentially raise new research questions. However, the science system can also develop in relation to problems arising in industrial contexts. New possibilities to patent arise unexpectedly as externalities within the research process, whereas in other (e.g., industrial) contexts scientific progress can sometimes be considered as an unintended by-product when the focus was initially on the solution of production problems (Rosenberg, 1990). The interactive and non-linear dynamics in the development of science, technology, and innovation also change professional practices. The new constellations drive the knowledge-based reconstructions of the political economies.[2]

---

[2] Reflexive mechanisms have increasingly been institutionalized in advanced industrial systems since the scientific-technical revolution of the period 1870-1910 (Braverman, 1974; Noble, 1977). In the first stage, the reconstructions remained confined to the physical and chemical properties of materials in the environment. More recently, this development has been reflected in sciences that reconstruct biological and institutional systems (Fukuyama, 2002).

## The study of knowledge-based communications

Knowledge-based systems do not exist in terms of stable elements, but they develop in terms of operations. Operations, however, can be combined and recombined in a variety of ways. As noted, several authors (e.g., Lundvall, 1992; Nelson, 1993) have proposed considering 'national systems of innovation' as the appropriate units of analysis for innovation studies. The choice of this national perspective allows for a direct link to the possibilities and limitations of policy making by national governments, and it enables the researcher to use national statistics (Lundvall, 1988). From a reflexive angle, however, an assumption about a (national) communality in the data remains a hypothesis (Skolnikoff, 1993; Andersen, 1994).

For example, the notion of a national identity is nowadays historically changing from a European perspective. The subnational construction of regions has resounded in some regions because of linguistic identities (e.g., Flanders, Catalonia), but in other places (as in France) regional authorities had to be shaped in order to accommodate European policies and harmonization. In other words, the units of analysis and the systems of reference can be considered as constructs that provide the analysis with a heuristics. What is relevant from one perspective can be considered as contextual from another. Innovations and knowledge-based reconstructions occur by definition at interfaces and therefore allow for more than a single angle of theoretical appreciation. Consequently, the categories in this reflexive field of studies have to be entertained reflexively, that is, as hypotheses.

A second argument against reifying one's categories naturalistically follows from a reflection on how to declare the time axis in the research design. In contrast to a historical build up, the evolutionary dynamics continuously operate in the present and with hindsight, that is, upon the instantiations of the systems under study (Giddens, 1984). The addition of a virtual dimension to the system at the Internet highlights these evolutionary dynamics. The global dimension tends to invert the time axis in the analysis by reconstructing the past from the perspective of other possibilities perceived in the present or more recently. Note that this development is only a tendency, since the global developments remain embedded in historical ones. However, the retrospective view provides us with an analytical angle for the construction of alternatives that is knowledge-based, since it is no longer limited by what was already constructed previously.

For example, an analysis of the strengths and weaknesses of a research portfolio does not by itself suggest that one should 'pick the winners' in order to strengthen one's case globally, that is, at the system's level (Irvine & Martin, 1984). The 'winners' may have been yesterday's winners and one may have other reasons to strengthen the hitherto relatively weak groupings or clusters (Porter, 1990). Empirical analyses inform us about the contingencies that can be expected; but since the dynamics are complex, unintended consequences and unforeseeable externalities are also expected (Callon, 1998). The evaluation provides us with signals that can sometimes be made the subject of systematic analysis by taking another angle.

# Operationalization

How can one move from the analysis of knowledge-based systems to a determination of the relative importance of the theoretical concepts in explaining an observable reality? How can a reflexive analyst make a convincing argument when the notion of a system of reference can always be deconstructed, and the time line may also be inverted in terms of what the historical accounts mean for the present?

Since systems that contain and communicate knowledge cannot be considered as given or immediately available for observation, one has to specify them analytically—that is, on theoretical grounds—before they can be indicated and/or measured. To this end the quantitative measurement remains thoroughly dependent on qualitative understanding. For example, one can raise the question of whether 'Mode 2' currently prevails in the production of scientific knowledge. What would count as a demonstration of this prevalence, and what as a counterargument? Can, for example, instances be specified in which one would also be able to observe processes of transition between the two modes? What should one measure in such instances, and why?

While qualitative analysis reduces the complexity by taking a perspective, quantitative analysis allows us to raise questions about the extent to which a theoretical perspective highlights a relevant dimension. Can the current development of 'biotechnology' in Germany be characterized as 'Mode 2'? How can it be compared with 'biotechnology' in the United States? (e.g., Giesecke, 2000) A policy analyst may always be able to point to contingency, sameness and differences, continuities and change, but the quantitative analysis requires that these categories be specified as *ex ante* hypotheses so that the expectations can be updated by the research efforts.

Empirical research enables us to specify the percentage of the variation that can be explained using one model or another. Whether 'Mode 2' is 'old wine in new bottles' (Weingart, 1997) or new wine in old bottles depends on the definitions of the bottles and the wines, and the processes of change that are analytically explicated. In other words, the definitions of a knowledge-intensive system are themselves knowledge-intensive (Nowotny *et al.*, 2001; Leydesdorff, 2001b). The observations and indicators are also knowledge-intensive, since one can no longer assume that the overwhelmingly available information would answer the research questions precisely (Hicks & Katz, 1996). A crucial question becomes the theoretically informed specification of a selection from the data. Which are the proper systems of reference?

For example, what one understands nowadays under the name of 'biotechnology' is very different from what governments wanted to stimulate in the 1980s (Nederhof, 1988). Analogously, what industries subsume under the category of 'biotechnology' is different from what research councils indicate with this same term. A modern society is pluriform and therefore differentiated in terms of its coordination mechanisms, codifications, and media of communication. The evolutionary perspective then demands an *ex post* delineation of the domains under study, but in the form of proposals and hypotheses.

Thus, the indication of knowledge-based systems is based both on theoretical reflection and on methodological considerations about how one may be able to proceed from the specific choices to operationalization, and vice versa. For example, one can operationalize

'biotechnology' in terms of a set of biotechnology journals in the *Science Citation Index*. If one fixes this journal set *ex ante* in order to make comparisons along the time line possible (Narin, 1976; Irvine *et al.*, 1985), one observes the development of 'biotechnology' as conceptualized at the beginning of the data collection. If one defines the journal set dynamically, one studies the changing meaning of 'biotechnology' in relation to other journal clusters. If one determines the journal set *ex post* one refers to the most recently available understanding. The latter definition can be made relevant for policy, while the former definitions inform historical studies (Leydesdorff, 2002).

In other words, the operationalization remains thoroughly dependent on the theoretical perspective. One looses the notion of independence of the external referent when studying knowledge-based systems from an evolutionary perspective (Luhmann, 2002). For example, using a journal set provides us with a focus on the scientific publication system. The use of patent data provides us with a focus on technological inventions. These two systems are differently codified and therefore can be expected to exhibit different dynamics. The methodological problems reflect decisions that have to be taken on analytical grounds. The theoretical considerations, however, can only be made relevant for the measurement if they can be formulated as hypotheses that are to be operationalized.

## Patent indicators

In order to demonstrate my point, let me provide data based on the U.S. national patent database, on the one hand, and based on the Internet as a globally developing system, on the other. 'University', 'industry', and 'government', and the various combinations with Boolean 'AND' operators can be used as keywords in these databases. As above (see Figure 1), I searched the patent database for the number of occurrences of the terms in the file on a year-to-year basis. For reasons of comparison with the Internet searches (below), the time series is limited to the period 1993-2002. Table 1 first provides the results of these searches.

| year | University | Industry | Government | UI | UG | IG | UIG | Total number of patents |
|------|-----------|----------|------------|------|------|------|-----|-------------------------|
| 1993 | 3063 | 9716 | 2619 | 401 | 588 | 334 | 63 | 110540 |
| 1994 | 3359 | 10568 | 2855 | 479 | 684 | 390 | 89 | 114564 |
| 1995 | 3710 | 10800 | 2828 | 529 | 771 | 410 | 93 | 114864 |
| 1996 | 4552 | 12147 | 3149 | 703 | 963 | 488 | 114 | 122953 |
| 1997 | 5406 | 12699 | 3604 | 814 | 1199 | 583 | 168 | 125884 |
| 1998 | 7623 | 17068 | 4708 | 1254 | 1658 | 807 | 266 | 166801 |
| 1999 | 8326 | 18553 | 4856 | 1352 | 1735 | 844 | 235 | 170265 |
| 2000 | 8488 | 19368 | 4831 | 1399 | 1776 | 865 | 267 | 176350 |
| 2001 | 9190 | 20812 | 5136 | 1591 | 1868 | 996 | 296 | 184172 |
| 2002 | 9228 | 21089 | 5242 | 1619 | 1928 | 1047 | 352 | 184531 |

**Table 1**
The number of hits for the search terms 'university,' 'industry,' and 'government' and their combinations in the database of the U.S. Patent and Trade Office

Note that these results do not indicate intellectual property or institutional relationship. The values for 'university AND industry' (UI), 'university AND government' (UG), and 'industry

AND government' (IG) can be considered as indicators of the bilateral links in the discursive domain of the database, whereas the value of UIG represents the trilateral communality between these three concepts.

In general, this type of data enables us to span a three-dimensional model as exhibited in Figure 3.

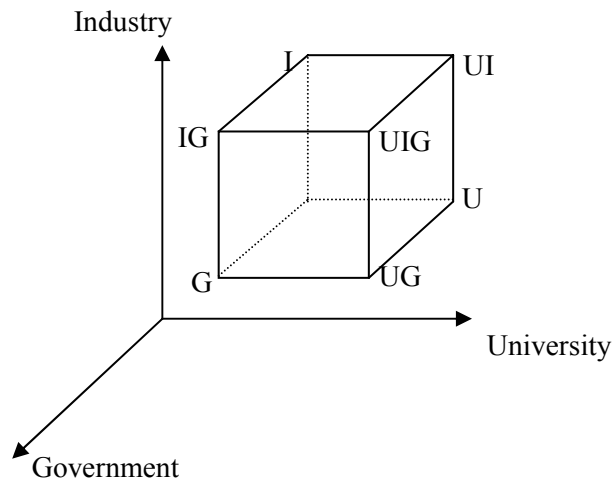

**Figure 3**
University-Industry-Government relations in three dimensions

As different from co-variation between two dimensions or co-occurrence measurement, mutual information or transmission can be defined analytically in three dimensions (Abramson, 1963).[3] Two states of a triple helix configuration can then be distinguished: in Figure 4 the three sets exhibit an overlap, whereas in Figure 5 this overlap has vanished. The mutual information in three dimensions ($T_{UIG}$) may become negative in the latter case, while this indicator has a positive value in the former.

---

[3] The transmission in three dimensions (x, y, z) can be defined as follows (Abramson, 1963, at p. 129):

$$T(xyz) = \Sigma_{xyz}\ P(xyz) \log \{[P(xy).P(xz).P(yz)] / [P(x).P(y).P(z).P(xyz)]\}$$

Or in another notation:

$$T(xyz) = H(x) + H(y) + H(z) - H(xy) - H(yz) - Hxz) + H(xyz)$$

In the first formulation, P(x) stands for the probability of an event x and P(xy) for the probability that x and y occur together, etc. These probabilities can be measured by counting frequencies of (co-occurrences) of events, as will be shown in the empirical examples below.

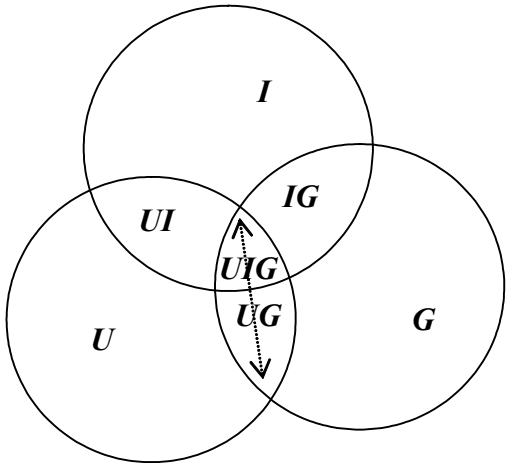 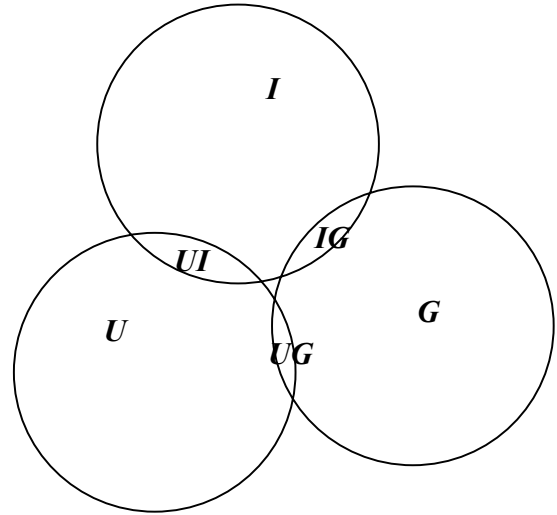

**Figure 4**
Three subsystems with a center of coordination

**Figure 5**
Three subsystems without a center of integration

When the three sets of documents containing the search terms 'university', 'industry', and 'government' are closely coupled by sharing a communality in the variation (e.g., in the case of neo-corporatist arrangements), the value of this transmission is positive. When the three subsets are completely uncoupled, the mutual information vanishes ($T_{UIG} = 0$). However, when the three domains are operationally coupled through uncoordinated bi-lateral relations, the indicator can also become negative. Thus, this indicator provides us with a measure for the state of a Triple Helix system whenever the dimensions can be specified so that the relevant relations can be counted.

Conceptually, the generation of a negative entropy such as mutual information corresponds with the idea of complexity that is contained or 'self-organized' in a network of relations that lacks central coordination. The system then propels itself in an evolutionary mode. The 'global' reduction of the uncertainty by the negative transmission is a result of the network structure of bi-lateral relations (Figure 6).

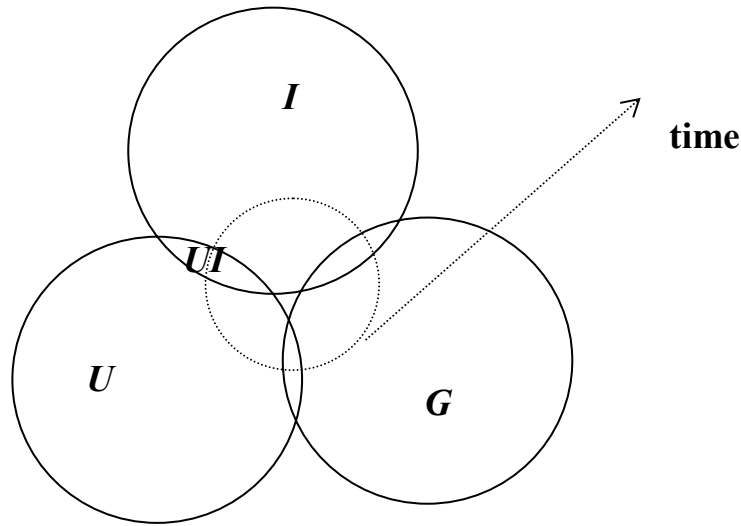

**Figure 6**
Three subsystems with hypercyclic integration in a globalized dimension

The next-order structure operates globally by constraining and enabling local substructures. However, the overall structure cannot be perceived completely from any of the positions in this networked system since there is no center of coordination. However, an evolving structure in a virtual dimension can be hypothesized and then also be attributed a value using the algorithm as an indicator. The globalizable expectations remain embedded in the local situations, albeit in a distributed and therefore uncertain mode. The embedded uncertainties cannot be observed, but by using an algorithmic indicator one can appreciate this latent structures of coordination.

Figure 7 provides the value of $T_{UIG}$ for the time-series of patent data during 1976-2002. The figures show that co-occurrences between two of the three terms prevail to the extent that the value of $T_{UIG}$ is negative, but that the discourse in the U.S. national patent system became further integrated in terms of making references to university-industry-government relations during the 1990s.

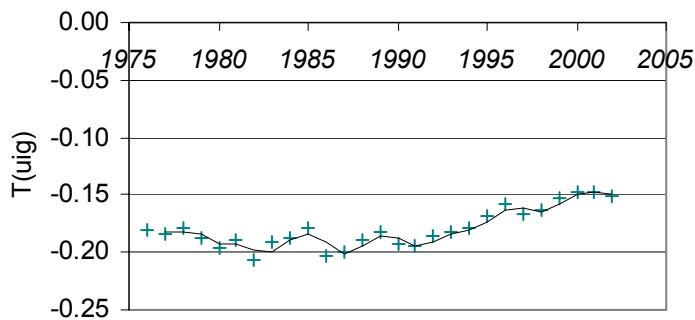

**Figure 7**
The mutual information among 'university,' 'industry,' and 'government' relations in the database of the U.S. Patent and Trade Office. (The curve added depicts the two-year moving averages.)

As noted, the Bayh-Dole Act can with hindsight be considered as having provided the patent system with one more degree of freedom, that is, by allowing universities increasingly to become players in this institutional field (Sampat *et al.*, 2003). The patent system, however, is a system of legal control by a national government and therefore under the pressure of integration. New players can be expected to be enrolled within this discourse, but it takes time to reshape the mutual perspectives.[4]

## Webometric data

Despite the poor operationalization of the industrial dimension when using the word 'industry' as a search term,[5] the increasing integration in the patent database is not a trivial result. This is demonstrated by the next test: Figure 8 is based on performing precisely the same exercise at the Internet using the *AltaVista Advanced Search Engine*.[6] In this case, the mutual information in three dimensions decreases during the second half of the 1990s.

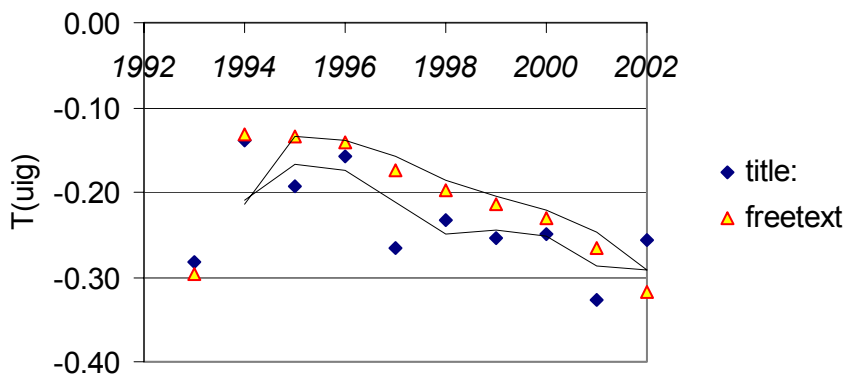

**Figure 8**
The mutual information among 'university,' 'industry,' and 'government' relations as retrieved at the Internet using the *Altavista Advanced Search Engine*. (The curves are based on two-year moving averages.)

In order to control for the effect of using these search terms without control, the searches were additionally conducted with the title-words. Title words are deliberately provided by the web-authors as meta-tags. Although the numbers of the retrievals are orders of magnitude smaller, the results are similar in exhibiting the trend. After a period of initial construction of the system (1992-1995), the value of the indicator decreases steadily. This 'self-organization' of the Triple Helix relations in the global dimension at the Internet seems to have flattened in the most recent years. Perhaps the flattening of the curve illustrates that the process of endogenous expansion of the Internet has been interrupted temporarily as the e-business has gone into a recession during these last years.

---

[4] During the period 1976-1992, $T_{UIG}$ had remained equal to $-0.190 \pm 0.008$.
[5] More than 50% of the patents contains an industrial address (Jaffe & Traitenberg, 2002), whereas only 10-20% are indicated under 'industry' in Table 1.
[6] I used the *AltaVista Advanced Search Engine* because this engine is unique in allowing searches with both Boolean operators and time delineations. For the methodological problems involved in using this tool, see (Leydesdorff, 2001a).

| year | University | Industry | Government | UI | UG | IG | UIG | "url:*" (total) |
|---|---|---|---|---|---|---|---|---|
| 1993 | 2205 | 441 | 1041 | 49 | 49 | 46 | 25 | 18437 |
| 1994 | 12722 | 2178 | 3579 | 1007 | 1174 | 719 | 391 | 135265 |
| 1995 | 66719 | 13190 | 21187 | 5140 | 6861 | 4541 | 2036 | 640967 |
| 1996 | 216548 | 45938 | 66839 | 16257 | 21729 | 15894 | 6945 | 2308162 |
| 1997 | 478164 | 110434 | 166550 | 37122 | 51259 | 35230 | 16224 | 5740624 |
| 1998 | 842665 | 243611 | 343066 | 71306 | 95478 | 78922 | 32318 | 14379504 |
| 1999 | 1415659 | 471387 | 669844 | 131979 | 178892 | 157446 | 61899 | 33053057 |
| 2000 | 3005285 | 975976 | 1385296 | 245470 | 342218 | 298731 | 117318 | 86537251 |
| 2001 | 5381142 | 2419632 | 3014141 | 523922 | 724722 | 679407 | 247734 | 186175482 |
| 2002 | 10408179 | 7779754 | 7301276 | 1216090 | 1646210 | 1567669 | 550263 | 492815972 |

**Table 2**
The number of hits for the free-text search terms 'university,' 'industry,' and/or 'government' and their combinations using the *AltaVista Advanced Search Engine* (May 15, 2003)[7]

Table 2 provides the data underlying this representation in a format similar to that of Table 1. As in the case of the patent data, the changes are not apparent by visual inspection of the data. Unlike variables, the study of fluxes requires an algorithmic approach and the results can therefore be counter-intuitive. Note that the Internet data are time-stamped in the present (in this case at May 15, 2003). As the Internet evolves, previous representations are continuously overwritten. The search engines also change, using additionally their own reflexive dynamics (Leydesdorff, 2001a).

While the words and title-words can be considered as variation, following the hyperlinks enables us to map the selections that the authors of the webpages make from the materials previously made available. In this dimension the authors can be expected to integrate references into their text, whereas they are expected to reach out using words and title words (Leydesdorff, 1989). The *AltaVista* search engine enables us to map these hyperlinks to the relevant domains in terms of their institutional affiliation using the extensions '.edu', '.com', and '.gov' as proxies. Note that these proxies are limited to the U.S.A. in the case of the .edu and .gov-domains, whereas the .com-domain is used worldwide.

The resulting figure (Figure 9) shows a mirror image to the curves exhibited in Figure 8. The selecting documents differentiate using their own codes in the present by using the selected documents as their knowledge-base. The knowledge base is thus integrated into the relatively stabilized instantiations taking part in the observable knowledge infrastructure (Giddens, 1984).

---

[7] The search "url:*" provided a total of 1,504,185,772 hits using the *Altavista* Advanced Search Engine on May 15, 2003.

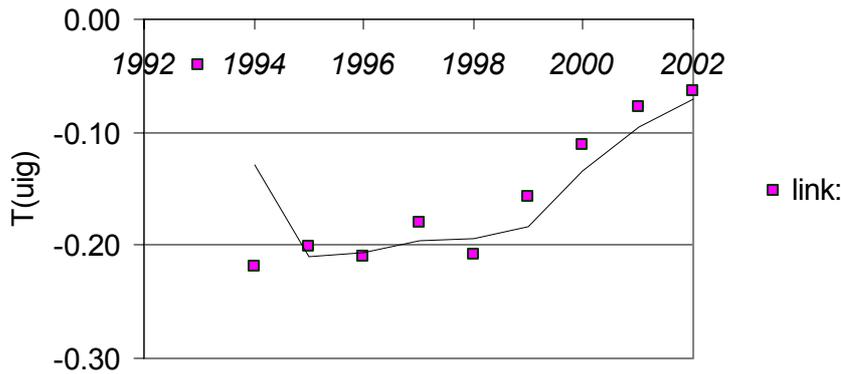

**Figure 9**
The mutual information among 'link:.edu,' 'link:.com,' and 'link:gov' relations as retrieved at the Internet using the *Altavista Advanced Search Engine* (15 May 2003).

| year | .edu | .com | .gov | edu AND com | edu AND gov | com AND gov | edu AND com AND gov | Link:* (total) |
|---|---|---|---|---|---|---|---|---|
| 1993 | 721 | 753 | 26 | 32 | 16 | 21 | 13 | 140631 |
| 1994 | 10653 | 5969 | 5070 | 1281 | 454 | 1657 | 264 | 155429 |
| 1995 | 58559 | 85344 | 63208 | 16060 | 4168 | 30666 | 2707 | 971806 |
| 1996 | 185571 | 213755 | 40505 | 52853 | 13816 | 15191 | 9713 | 4215445 |
| 1997 | 383999 | 586804 | 76767 | 118249 | 25447 | 29842 | 18723 | 8410235 |
| 1998 | 714592 | 1512795 | 206683 | 177352 | 49238 | 59734 | 33695 | 21190676 |
| 1999 | 1410789 | 3372441 | 341635 | 346610 | 92354 | 126961 | 63192 | 42521722 |
| 2000 | 2212642 | 10057844 | 577433 | 622780 | 194573 | 244278 | 151641 | 92177426 |
| 2001 | 3722856 | 30497559 | 1328142 | 1344270 | 373437 | 599161 | 305180 | 196204140 |
| 2002 | 8564790 | 81698935 | 4035084 | 3058198 | 1159347 | 1758589 | 757120 | 501734312 |

**Table 3**
The number of hits for the search terms 'link:university,' 'link:industry,' and/or 'link:government' and their combinations using the *AltaVista Advanced Search Engine* (May 15, 2003)

Table 3 shows the number of links involved. The number of links with .com is an order of magnitude larger than the other ones and among the bilateral links the co-'sitation' of .edu and .com by the referring documents prevails (Rousseau, 1997).[8] Although the more skewed shape of these distributions considerably reduces the values of the entropies involved, the interaction effect among the entropies is not visible on the basis of the values of the different fluxes without further computation.

---

[8] The number of links to .com pages is also an order of magnitude larger than the number of hits using 'industry' as a free text term (Table 2).

## The measurement of complex and codified communications

What can the above pictures teach us? As noted, the word 'university' can be expected to mean something different in a patent application than on the Internet. Furthermore, the meaning of a word may change over time. For example, it may have become more important for an applicant to make his or her collaboration with a university visible in a patent application without necessarily implying that these collaborations did not exist previously. A pervasive problem with measurement in the case of complex dynamics is that both the values of the variables and also the meanings of the variables may change with the choice of the system of reference and over time. If one tries to measure change in both the meanings of the variables and the values of the variables using a single design, the understanding tends to become confused because one loses a clear definition of a baseline (Studer & Chubin, 1980, at pp. 269 ff.).

Knowledge-based developments cannot be equated with the development of institutional units (Collins, 1985) or with fixed journal sets (Narin, 1976). The evolutionary focus on flows of communications makes it necessary first to hypothesize *what* each system of communications is communicating when it operates. For example, a system of references (citations, outlinks) can be expected to communicate differently from a system of co-occurrences of words or a (re-)distribution of institutional addresses. Citations relate papers along trajectories over time, whereas institutional addresses of coauthorships, for example, can also be used for the mapping at specific moments in time.

The specification of a system of reference in terms of an *operation*—as different from a unit of analysis—extends the analysis with a reflection on the time horizon. In the historical dimension, I have elaborated above on the issue of inverting the time axis because statisticians have been inclined to build on databases using a historical perspective. Historically interested sociologists and socio-constructivists share this interest in temporal order in the materials: the quantitative data can then be used mainly as illustrations for the narratives when 'following the actors' (Latour, 1987). The study of knowledge-intensive developments, however, requires us to take a reflexive turn towards the data gathering process, both in the quantitative and in the qualitative domains. The focus is no longer on the actors, but on the emerging order in their communications (Urry, 2000; Leydesdorff, 2002).

## Conclusion

I have argued that a fundamental reformulation of the problems of science, technology, and innovation policies became urgent during the 1990s. Three models have been proposed for the study of innovation systems: (i) the distinction of a 'Mode 2' type of knowledge production, (ii) the model of 'national systems of innovation,' and (iii) the Triple Helix model of university-industry-government relations.

The authors of the 'Mode 2' thesis (Gibbons et al., 1994) have argued that the new configuration has led to a de-differentiation of the relations between science, technology, and society. Internal codification mechanisms (like 'truth-finding') were discarded by these authors as an 'objectivity trap' (Nowotny et al., 2001, at pp. 115 ff.). From this perspective, all scientific and technical communication can be translated and compared with other communication from the perspective of science, technology, and innovation policies (Callon et al., 1986; Latour, 1987).

In my opinion, the 'Mode 2' model focused on the performative integration of representations of systems that are otherwise different and continuously also differentiating. The systems under study are asymmetrically integrated at the historical interfaces, for example, in the case of successful innovations. However, they can be expected otherwise to restore their own orders by differentiating again in terms of the specificity of their respective communication codes. This asymmetry of the differentiation is needed in order to perform a next cycle of integration.

Differentiation and integration do not exclude one another, but rather depend on one another as different dimensions of the communication over time. A specific integration can be expected to mean something different in the various dimensions that were integrated. The communication enables us to construct and sometimes stabilize an integration, but the underlying systems compete both in terms of their definitions of social realities and in terms of the representations that they construct at the localizable interfaces. Systems of innovations solve the puzzles of how to interface different functions in the communication. The solutions and failures are manifest at the level of historical organization. The latter can then also be reshaped.

Evolutionary economists have argued in favor of studying 'national systems of innovation' as hitherto the most relevant level of integration. Indeed, they have provided strong arguments for this choice (Lundvall, 1992; Nelson, 1993; Skolnikoff, 1993). However, these systems are continuously being restructured under the drive of a global differentiation of expectations. Economies are interwoven both at the level of markets and in terms of multinational corporations, sciences are organized internationally, and governance is no longer limited within national boundaries. The most interesting innovations can be expected to involve boundary-spanning mechanisms.

In sum, I concur with the 'Mode 2'-model in assuming a focus on communication as the driver of systems of knowledge production and control. However, the problem of structural differences among the communications and the organization of interfaces remains crucial to the understanding of innovation in a global and knowledge-based economy. The wealth of knowledge and options for further developments have to be retained by reorganizing institutional arrangements with reference to global horizons.

The Triple Helix model tries to capture both dynamics by introducing the notion of an overlay that feeds back on the institutional arrangements. Each of the helices develops internally, but they also interact in terms of exchanges of both goods and services and in terms of knowledge-based expectations. The various dynamics have first to be distinguished and operationalized, and then sometimes they can also be measured. I have argued that the dynamics among the dimensions can then be measured using algorithmic indicators.

The strength of this research program is that it is no longer assumed to be possible to generalize on the basis of intuitions and naturalistic assumptions about the data. Empirical results are expected to inform us, but the results can also be counterintuitive. One may be able to appreciate unexpected results by innovating one's theoretical assumptions about the relevant systems of reference. If the various subdynamics can be better understood, one may also be able to develop simulation models on the basis of the reconstructions.

There is an intimate connection between the algorithmic evaluation of indicators and simulation studies. When analyzing knowledge-based systems, (scientometric) indicators enable us to study knowledge production and communication in terms of the traces that communications leave behind, while simulations try to capture the operations and their possible interactions. The common assumption is that knowledge production, communication, and control are considered as operations that change the materials on which they operate. The historically observable units of analysis are reflexively supplemented with units of operation that can only be specified on the basis of theoretical assumptions. Because of this dependency on the theoretical specification for its measurement, a knowledge-based economy can be expected to reinforce its development.